\documentclass{elsarticle}
\RequirePackage{packages}
\title{Deterministic and Probabilistic Rounding Error Analysis for Mixed-Precision Arithmetic on Modern Computing Units}
\author{Sahil Bhola\corref{cor1}, Karthik Duraisamy}
\address{Department of Aerospace Engineering \& Michigan Institute for Computational Discovery and Engineering, \\University of Michigan, Ann Arbor, U.S.A.}
\date{}
\makeatletter
\def\ps@pprintTitle{%
 \let\@oddhead\@empty
 \let\@evenhead\@empty
 \let\@oddfoot\@empty
 \let\@evenfoot\@empty}
\makeatother
\begin{document}
\begin{abstract}
    Modern computer architectures support low-precision arithmetic, which present opportunities for the adoption of mixed-precision algorithms to achieve high computational throughput and reduce energy consumption.
    As a growing number of scientific computations leverage specialized hardware accelerators, the risk of rounding errors increases, potentially compromising the reliability of models.
    This shift towards hardware-optimized, low-precision computations highlights the importance of rounding error analysis to ensure that performance gains do not come at the expense of accuracy, especially in high-stakes scientific applications.
    In this work, we conduct rounding error analysis on widely used operations such as fused multiply-add (FMA), mixed-precision FMA (MPFMA), and NVIDIA Tensor cores.
    We present a deterministic and probabilistic approach to quantifying the accumulated rounding errors.
    Numerical experiments are presented to perform the multiply and accumulate operation (MAC) and matrix-matrix multiplication using Tensor cores with random data.
    We show that probabilistic bounds produce tighter estimates by nearly an order of magnitude compared to deterministic ones for matrix-matrix multiplication.
\end{abstract}
\maketitle
\section{Introduction}\label{sec:introduction}
% Modern heterogeneous architecture provides dedicated hardware to accelerate specific tasks.
% For example, NVIDIA's Tensor Cores (TCs) and Google's matrix units (MXUs) are dedicated to accelerate matrix-matrix multiplication, a fundamental kernel in linear algebra.
% These hardware typically leverage low- or mixed-precision arithmetic to reduce the energy footprint and memory bandwidth consumption in large-scale applications.

Rapidly growing computational demands of artificial intelligence workloads have fundamentally reshaped hardware architectures, introducing specialized accelerators and novel arithmetic formats optimized for neural network operations. This transformation of the hardware landscape requires a corresponding evolution in scientific computing paradigms to effectively utilize these architectures.
Double precision $\float{64}$ has been the default choice for applications in science and engineering.
Recently, an increasing number of applications such as deep learning~\cite{gupta2015deep,wang2018training,hubara2018quantized}, climate modeling~\cite{paxton2022climate,kimpson2023climate,hatfield2019accelerating}, solution of linear systems of equations~\cite{higham2019squeezing,haidar2020mixed,haidar2017investigating,abdelfattah2021survey,buttari2007mixed}, fluid dynamics~\cite{liu2004real,rinaldi2012lattice,karp2023uncertainty}, and natural sciences~\cite{lienhart2002using} have begun leveraging low- or mixed-precision arithmetic for better performance.
However, with improved computational performance comes a dramatic drop in computational accuracy as rounding errors become prominent with low-precision arithmetic.
This warrants that any application leveraging low- or mixed-precision arithmetic must be aware of the accumulated rounding error to ensure that performance gains are not obtained at the expense of accuracy.
% Thus, motivating a rigorous rounding error analysis.

Traditional deterministic rounding error analysis (DBEA) assumes the worst-case scenario, where the absolute rounding errors are identical to the unit roundoff $\urd$~\cite{higham2002accuracy}.
This approach produces error bounds that depend on the constant $\gamma_{n}\define {n\urd} / (1 - n\urd)$ that holds for $n\urd < 1$, where $n$ is the number of arithmetic operations.
Despite providing strong guarantees, DBEA dramatically overestimates the error for large-scale problems using low-precision arithmetic.
This results in a conservative selection of precision.
Probabilistic treatment of rounding error can account for the potential cancellation and magnification of rounding errors and therefore provide better estimates of the accumulated rounding error~\cite{von1947numerical,hull1966tests,henrici1966test,tienari1970statistical,ipsen2020probabilistic}.
Previous studies~\cite{higham2020sharper,bhola2024variance} have shown that by treating the rounding error as independent and identically distributed (i.i.d.) random variables, we can obtain error bounds that depend on the constant $\tilde{\gamma}_n \propto \sqrt{n}~\forall n$.
As a result, probabilistic models scale well with problem size and can, therefore, provide better estimates of the accumulated rounding errors for large-scale problems.

This work examines the fused-multiply and accumulate (FMA), mixed-precision fused-multiply and accumulate (MPFMA), and matrix-matrix multiplication using Tensor Cores (TCs).
These kernels are widely used in numerical linear algebra as building blocks for kernels such as dot-products, block LU factorization, and Cholesky factorization.
We present both the deterministic and probabilistic rounding analysis for these kernels.

The manuscript is organized as follows.
In~\cref{sec:fp}, the rounding error due to floating-point representation and its interpretation using backward and forward errors is presented.
In~\cref{sec:kernels}, we perform deterministic and probabilistic rounding error analysis for FMA, MPFMA, and matrix-matrix multiplication using TCs.
In~\cref{sec:results}, the numerical experiments are presented for the multiply and accumulate operation and performing matrix-matrix multiplication.
In~\cref{sec:conclusions}, we present the concluding remarks for this work and provide future directions.

\section{Floating-point Arithmetic}\label{sec:fp}
\begin{figure}
    \centering
    \includegraphics[scale=0.4]{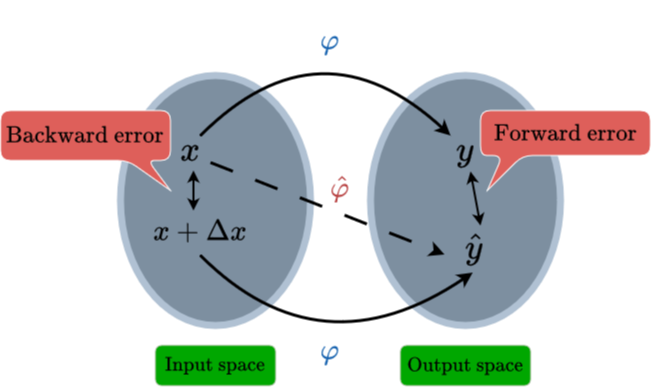}
    \caption{Schematic of backward and forward errors. Solid line: exact; dashed line: computed.}
    \label{fig:fwd_bwd_schematic}
\end{figure}

For a given \textit{precision} $p\in\mathbb{N}$, \textit{base} $\beta$, and \textit{exponent range} $e_{min} \leq e \leq e_{max}$, we can define the floating point number system $\mathbb{F}\subset \real{}$ as
\begin{align}
    \mathbb{F} \define \Big\{ (-1)^s d_0.d_1 d_2\dots d_{p-1} \times \beta^e \Big| s\in\{0, 1\}, 0 \leq d_i \leq \beta - 1 \Big\},
\end{align}
where $d_0 \neq 0$ for normalized numbers.
Given the tuple $t^*\define (p, \beta, e_{\min}, e_{\max})$, a finite and unique set of \textit{representable} numbers can be obtained that constitute the floating-point system.
Standard floating-point systems include IEEE half (\float{16}), single (\float{32}), and double (\float{64}) precision, for which the tuple parameters are tabulated in~\cref{tab:fp_params}.
Thus, representing $z\ireal{}$ in a floating-point system requires a \textit{rounding operation} $fl: \real{} \rightarrow \mathbb{F}$ that maps $z$ to a floating-point representable number according to a rounding model.
% Typically used rounding models include round-to-nearest, stochastic-rounding, and round towards/away from zero.
Due to this rounding operation, there is a loss of information (or bits) that introduces a \textit{rounding error}, which can grow with successive floating point operations.
Thus, computing any real-valued function $\varphi: \real{n_{in}} \rightarrow \real{n_{out}}$ in finite-precision produces an approximate $\hat{\varphi}: \mathbb{F}^{n_{in}}\rightarrow \mathbb{F}^{n_{out}}$, as illustrated in~\cref{fig:fwd_bwd_schematic}.
To quantify the approximation quality, we can then define \textit{forward error} as the absolute or relative error specified in the output space.
Alternatively, the effect of rounding can be understood as perturbations to the input space that propagate through the true function.
To this end, we can define a perturbation $\Delta \vect{x}$ to the input space $\vect{x}$, such that $\varphi(\vect{x} + \Delta \vect{x}) = \hat{\varphi}(\vect{x})$.
Since the input perturbation can be non-unique, we can define the smallest relative perturbation called the \textit{backward error} as the solution of the optimization problem
\begin{equation}
    \epsilon_{bwd} \define \min \{\epsilon\geq 0: \hat{\varphi}(\vect{x}) = \varphi(\vect{x} + \Delta \vect{x}), \abs{\Delta \vect{x}} \leq \epsilon \abs{\vect{x}}\}.
\end{equation}

Forward or backward error bounds can be developed to quantify the rounding uncertainty.
In this work, we focus on developing backward error bounds, which can be directly compared with uncertainties in the input space.
This is useful because for a given floating-point system parameterized by $ t^*$, if the inherent uncertainty in the input space is larger than the backward error, then the computed solution can be considered reliable.
Further, backward error bounds can also be used to develop forward error bounds that are useful for uncertainty propagation.

\subsection{Rounding Error Analysis}\label{sec:rounding}
The rounding uncertainty resulting from finite-precision approximation implicitly depends on the rounding operations.
Using the IEEE standard~\cite{ieee2019ieee}, the rounding operation can be modeled as
\begin{align}
    fl(z) \define z (1 + \delta)^{\rho};\quad z\ireal{}, \rho=\pm 1, \abs{\delta} \leq \urd
\end{align}
where $\delta$ is the \textit{rounding error}, and $\urd \define \frac{1}{2}\beta^{1-p}$ is the \textit{unit roundoff}.
Thus, the rounding error terms accumulate when computing the true function $\varphi$ in finite precision.
For example, evaluating $y = \sum_{i=1}^3 x_i $ in finite precision results in an approximation $\hat{y} = \sum_{i=1}^3 x_i (1 + \eta_i)\prod_{j=\max(2, i)}^{3} (1 + \delta_j)$, assuming recursive summation from left-to-right.
Here, $\eta_i$ is due to the rounding $fl(x_i)$, and $\delta_i$ arises due to the rounding after the addition operation.
In backward error analysis, a bound for the product $\prod_{i=1}^{n} (1+\delta_i)$ is sought to obtain a perturbed system.
\paragraph{Deterministic Backward Error Analysis (DBEA)} In the traditional backward analysis, the following lemma~\cite[Lemma 3.1]{higham2002accuracy} is used to bound the distance between $1$ and the product $\prod_{i=1}^n(1+\delta_i)^{\rho_i}$.
\begin{lemma}[Deterministic rounding error bound]
\label{lemma:dbea}
If $\abs{\delta_i}\leq \urd$ and $\rho_i=\pm 1$ for $i=1,\dots,n$, and $n\urd < 1$, then
\begin{equation*}
    \prod_{i=1}^n ( 1 + \delta_i)^{\rho_i} \define 1 + \theta_n^\urd,
\end{equation*}
with $\abs{\theta_n^\urd} \leq \gamma_n^\urd \define \frac{n\urd}{1-n\urd}$.
\label{lemma:dbea}
\end{lemma}

\paragraph{Variance-informed Probabilistic Backward Error Analysis (VIBEA)}
While the traditional deterministic analysis provides strong guarantees for the rounding uncertainty, it considers the worst-case scenario where the absolute rounding error is identical to the unit roundoff.
This dramatically overestimates the accumulated rounding errors, producing very pessimistic error bounds —especially when using low-precision arithmetic for a large number of operations.
To address this,~\cite[Theorem 3.4]{bhola2024variance} proposed the following lemma.
\begin{lemma}[Variance-informed probabilistic error bounds]
\label{lemma:vibea}
Let $\delta_1,\dots, \delta_n$ represent $n$ i.i.d. random variables distributed uniformly as $\mathcal{U}[-\urd, \urd]$ where $\urd \leq 1$.
Then for $\lambda \geq 0$
\begin{equation*}
    \prod_{i=1}^n (1+\delta_i)^{\rho_i} = 1 + \tilde{\theta}_n^\urd,
\end{equation*}
    hold such that $\abs{\tilde{\theta}_n^\urd} \leq \tilde{\gamma}_n^\urd(\lambda)$ with probability at least
\begin{equation*}
    \prob[b]{\lambda, \urd, n} \define 1 - 2\exp \Big ( \frac{-\lambda^2n\urd^2}{2(\sigma^2 + \frac{\lambda\sqrt{n}\urd^2}{3(1-\urd)})}\Big),
\end{equation*}
where
\begin{equation*}
    \sigma^2 = n \Big(\frac{4\urd^2 + \kappa(\log(1-\urd)^2 - 2\log(1-\urd)\log(1+\urd) + \log(1+\urd)^2)}{4\urd^2}\Big),
\end{equation*}
with $\kappa = (-1+\urd^2)$.
\end{lemma}

\begin{table}[ht]
\centering
\begin{tabular}{l c c c c c}
\toprule
    \textbf{Arithmetic} & $\mathbf{p}$ & $\mathbf{\beta}$ & $\mathbf{e_{min}}$ & $\mathbf{e_{max}}$ \\
\midrule
    IEEE half (\float{16})   & 11 & 2 & -14  & 15 \\
    IEEE single (\float{32}) & 24 & 2 & -126 & 127\\
    IEEE double (\float{64}) & 53 & 2 & -1022 & 1023 \\
\bottomrule
\end{tabular}
\caption{Floating point arithmetic parameters for IEEE standard.}
\label{tab:fp_params}
\end{table}

\section{Application to Numerical Linear Algebra}\label{sec:kernels}
In this section we first present the backward error bounds for the fused multiply-add (FMA) operation, mixed-precision FMA (MPFMA) operation.
Followed by, we extend the MPFMA backward error bounds to obtain the backward-forward error bounds for the NVIDIA Tensor Core operations.
\subsection{Fused Multiply-Add (FMA)}\label{sec:fma}
The FMA operation is a hardware-level instruction defined as $fma(a, b, c)$, computed as the correctly rounded exact result $fl_\urd(a\times b + c)$.
Due to fewer rounding operations and, thereby, smaller rounding errors, these operations are widely used in numerical algorithms such as computing dot products, generalized matrix-matrix multiplications (GEMM), and polynomial evaluations.
For example, computing the dot-product $z=\sum_{i=1}^{n} x_i y_i$ via recursive summation requires $n$ multiplications and $n-1$ summations.
Alternatively, we can compute the dot-product as a recursive FMA operation $s_i = fma(x_i, y_i, s_{i-1})$ for $i=1\dots n$, where $s_0=0$ and $z=s_n$.
In this section, we study the backward error bounds for the FMA operation.
\subsubsection{Deterministic Analysis}\label{sec:dbea_fma}
\begin{lemma}[FMA without representation error]
\label{lemma:dbea_fma_wo_rep}
Let the multiply-accumulate operation $z = a \times b + c$ be evaluated using the FMA operation, where $a, b, c \in\mathbb{F}$ characterized by the tuple $t^*$ with a unit roundoff $\urd$.
    Under the IEEE arithmetic model and~\cref{lemma:dbea}, the computed approximation $\hat{z}$ satisfies
\begin{align*}
\hat{z} = a\times (b+\Delta b) + (c+\Delta c) = (a+\Delta a)\times b + (c+\Delta c),
\end{align*}
where $\abs{\Delta a} \leq \urd \abs{a}$, $\abs{\Delta b} \leq \urd \abs{b}$, and $\abs{\Delta c} \leq \urd \abs{c}$.
\begin{proof}
The approximation $\hat{z}$ is the correctly rounding exact result given as
\begin{align*}
\hat{z} = fma(a, b, c) (1 + \delta) = (a \times b + c)(1 + \delta) ;\quad \abs{\delta} \leq \urd,\\
= (a\times b)(1 + \delta) + c(1 + \delta),\\
= a\times (b + \Delta b) + (c + \Delta c)= (a + \Delta a) \times b + (c + \Delta c),
\end{align*}
where $\abs{\Delta a} \define \abs{a \delta} \leq \urd \abs{a}$, $\abs{\Delta b} \define \abs{b \delta} \leq \urd \abs{b}$, and $\abs{\Delta c} \define \abs{c \delta} \leq \urd \abs{c}$.
\end{proof}
\end{lemma}
Using~\cref{lemma:dbea_fma_wo_rep}, we can then obtain the relative forward error bound as

\begin{align}
\frac{\abs{\hat{z} - z}}{\abs{z}} \leq \urd \frac{\abs{a}\times\abs{b} + \abs{c}}{\abs{a\times b + c}}.
\end{align}
If $a, b$, and $c$ are described by the tuple $t^{*}$ with a unit roundoff $\acute{\urd} < \urd$, where $\urd$ is the unit roundoff for the FMA operation, then we must account for the initial change in precision.
Using~\cref{lemma:dbea}
\begin{align*}
    \tilde{z} &= a \times (b + \Delta b) + c + \Delta c,\quad \abs{\Delta b} \leq \gamma_2^\urd\abs{b}; \abs{\Delta c} \leq \urd \abs{c},
\end{align*}
which is the expression that is then computed using the FMA operation.
Using~\cref{lemma:dbea_fma_wo_rep}, the computed approximation satisfies
\begin{align*}
    \hat{z} &= a \times (b + \Delta b + \Delta (b + \Delta b)) + (c + \Delta c) + \Delta (c + \Delta c),\\
    &= a \times b + c + a\Delta b + a\Delta (b + \Delta b) + \Delta c + \Delta (c + \Delta c),\\
    &\define a \times b + c + \Delta p,
\end{align*}
where $\abs{\Delta (b + \Delta b)} \leq \urd \abs{b + \Delta b}$ and $\abs{\Delta (c + \Delta c)} \leq \urd \abs{c + \Delta c}$.
We can then bound $\Delta p$ to obtain the forward error bound as stated in~\cref{lemma:dbea_fma_w_rep}.
\begin{lemma}[FMA with representation error]
\label{lemma:dbea_fma_w_rep}
    Let the multiply-accumulate operation $z = a \times b + c$ be evaluated using the FMA operation with a unit roundoff $\urd$, where $a, b, c \in \mathbb{F}$ characterized by the tuple $t^{*}$ with a unit roundoff $\acute{\urd} < \urd$.
    Under the IEEE arithmetic model and~\cref{lemma:dbea}, the computed approximation $\hat{z}$ satisfies
    \begin{align*}
        \frac{\abs{\hat{z} - z}}{\abs{z}} \leq \frac{(\urd + \gamma_2^\urd + \urd \gamma_2^\urd) \abs{a}\times\abs{b} + (2\urd + \urd^2)\abs{c}}{\abs{a\times b + c}}.
    \end{align*}
\end{lemma}

\subsubsection{Probabilistic Analysis}\label{sec:vibea_fma}
It is trivial to show that for an FMA operation without representation error, DBEA and VIBEA result in the same backward error bound, which always holds.
In the case there is an initial change in precision, we can use~\cref{lemma:vibea} to obtain
\begin{align*}
    \tilde{z} = a \times (b + \Delta b) + c + \Delta c,
\end{align*}
where $\abs{\Delta b} \leq \tilde{\gamma}_2^\urd\abs{b}$ holds with probability at least $\prob[b]{\lambda, \urd, 2}$ and $\abs{\Delta c} \leq \urd \abs{c}$.
Following~\cref{sec:dbea_fma}, the probabilistic forward error bounds can be obtained as stated in~\cref{lemma:vibea_fma_w_rep}.
\begin{lemma}[FMA with representation error]
\label{lemma:vibea_fma_w_rep}
    Let the multiply-accumulate operation $z = a \times b + c$ be evaluated using the FMA operation with a unit roundoff $\urd$, where $a, b, c \in \mathbb{F}$ characterized by the tuple $t^{*}$ with a unit roundoff $\acute{\urd} < \urd$.
    Under the IEEE arithmetic model and~\cref{lemma:vibea}, the computed approximation $\hat{z}$ satisfies
    \begin{align*}
        \frac{\abs{\hat{z} - z}}{\abs{z}} \leq \frac{(\urd + \tilde{\gamma}_2^\urd + \urd \tilde{\gamma}_2^\urd) \abs{a}\times\abs{b} + (2\urd + \urd^2)\abs{c}}{\abs{a\times b + c}},
    \end{align*}
    with a probability at least $\prob[b]{ \lambda, \urd, 2 }$.
\end{lemma}

\subsection{Mixed-precision FMA (MPFMA)}\label{sec:mpfma}
While the FMA operation presented in~\cref{sec:fma} delivers performance optimization via hardware-level instructions, it is typically performed with a fixed precision.
% This poses larger memory bandwidth requirement
Modern hardware provides a dictionary of lower precision, such as $\float{4}$ and $\float{8}$, that can be leveraged to reduce the memory bandwidth requirement and attain high throughput.
% The MPFMA performs the multiply-accumulate operation by taking in the inputs $a, b$ in a lower precision (with unit roundoff $\urd_{low}$) and $c$ in a higher precision (with unit roundoff $\urd_{high}<\urd_{low}$), as illustrated in~\cref{fig:mpfma}.
% The multiply operation takes place in
\begin{figure}[h]
    \centering
    \includegraphics[scale=0.3]{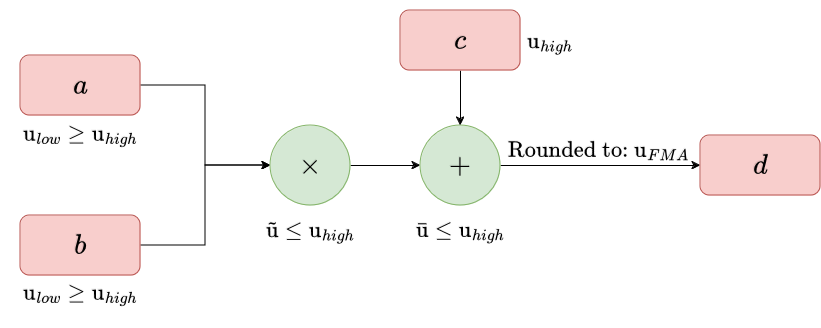}
    \caption{Schematic of the mixed-precision FMA operation.}
    \label{fig:mpfma}
\end{figure}
The typical MPFMA operation takes in the input $a$ and $b$ in a lower precision characterized by the tuple $t_{low}$ with a unit roundoff $\urd_{low}$, and $c$ in a higher precision characterized by the tuple $t_{high}$ with unit roundoff $\urd_{high}$.
This is illustrated in~\cref{fig:mpfma}.
The product $a \times b$ is performed in a higher precision characterized by the tuple $\tilde{t}$, with a unit roundoff $\tilde{\urd} \leq \urd_{high}$.
To ensure that this product does not introduce any rounding error, the low precision floating point system $t_{low}$ is chosen to represent the entire significant of the product in the floating point system $\tilde{t}$.
For example, the NVIDIA tensor cores utilize $\float{16}$ as the low precision and compute the product in $\float{32}$.
The product of two $\float{16}$ has at most $20$ significant bits, which can be entirely represented by the $23$ significant bits of $\float{32}$.
Following the product, the accumulation operation is performed in the floating point system $\bar{t}$ -- introducing a rounding error.
Lastly, the output is stored in a floating point system characterized by $t_{FMA}$ with a unit roundoff $\urd_{FMA}$.
If $\urd_{FMA} > \bar{\urd}$, then this step also introduces a rounding error.
In this section, we present the backward error analysis of the MPFMA operation.
\subsubsection{Deterministic Analysis}\label{sec:dbea_mpfma}
\begin{lemma}[MPFMA without representation error]
\label{lemma:dbea_mpfma_wo_rep}
    Let the multiply-accumulate operation $z = a\times b + c$ be evaluated using MPFMA operation, where $a b\in\mathbb{F}$ characterized by the tuple $t_{low}$ and $c\in\mathbb{F}$ characterized by the tuple $t_{high}$.
    Consider that the multiply and accumulate operation is performed in a floating point system characterized by $\tilde{t}$ (with unit roundoff $\tilde{u}$) and $\bar{t}$ (with unit roundoff $\bar{u}$), respectively.
    Let the output unit roundoff be given as $\urd_{FMA} > \bar{\urd}$.
    Then, under the IEEE arithmetic model and~\cref{lemma:dbea}, the computed approximation $\hat{z}$ to $z$ satisfies
    \begin{align*}
        \hat{z} = a \times (b + \Delta b) + (c + \Delta c) = (a + \Delta a) \times b + (c + \Delta c),
    \end{align*}
    where
    \begin{align*}
            \abs{\Delta a} \leq (\bar{\urd} + \urd_{FMA} + \bar{\urd}\urd_{FMA}) \abs{a},\\
            \abs{\Delta b} \leq (\bar{\urd} + \urd_{FMA} + \bar{\urd}\urd_{FMA}) \abs{b},\\
            \abs{\Delta c} \leq (\bar{\urd} + \urd_{FMA} + \bar{\urd}\urd_{FMA}) \abs{c}.
    \end{align*}
    \begin{proof}
        Assume that the significant of the product $a\times b$ can be entirely represented by the significant of the floating point system $\tilde{t}$.
        Then, the computed approximation is given as
        \begin{align*}
            \hat{z} &= (a\times b + c)(1 + \delta)(1 + \eta),\quad \abs{\delta} \leq \bar{\urd}, \abs{\eta} \leq \urd_{FMA},\\
            &= a \times (b + \Delta b)  + (c + \Delta c) = (a + \Delta a) \times b  + (c + \Delta c),
        \end{align*}
        where $\Delta a \define (\delta + \eta + \delta\eta) a$, $\Delta b \define (\delta + \eta + \delta\eta) b$ and $\Delta c \define (\delta + \eta + \delta\eta) c$.
        This gives
        \begin{align*}
            \abs{\Delta a} \leq (\bar{\urd} + \urd_{FMA} + \bar{\urd}\urd_{FMA}) \abs{a},\\
            \abs{\Delta b} \leq (\bar{\urd} + \urd_{FMA} + \bar{\urd}\urd_{FMA}) \abs{b},\\
            \abs{\Delta c} \leq (\bar{\urd} + \urd_{FMA} + \bar{\urd}\urd_{FMA}) \abs{c}.
        \end{align*}
    \end{proof}
\end{lemma}
Using~\cref{lemma:dbea_mpfma_wo_rep}, the relative forward error bound is given as
\begin{align}
    \frac{\abs{\hat{z} - z}}{\abs{z}} \leq (\bar{\urd} + \urd_{FMA} + \bar{\urd}\urd_{FMA})\frac{\abs{a}\times\abs{b} + \abs{c}}{\abs{a\times b + c}}.
\end{align}

In the special case where $\tilde{\urd} = \bar{\urd} = \urd_{FMA} = \urd_{high}$, there is only a single rounding operation, such that the computed approximation becomes
\begin{align*}
        \hat{z} = a \times (b + \Delta b) + (c + \Delta c) = (a + \Delta a) \times b + (c + \Delta c),
\end{align*}
where $\abs{\Delta b} \leq \urd_{high}$ and $\abs{\Delta c} \leq \urd_{high}$.
This results in a forward error bound
\begin{align}
    \frac{\abs{\hat{z} - z}}{\abs{z}} \leq \urd_{high}\frac{\abs{a}\times\abs{b} + \abs{c}}{\abs{a\times b + c}},
\end{align}
which is identical to the bound obtained when performing an FMA operation, as shown in~\cref{lemma:dbea_fma_wo_rep}.

If $a, b, c$ are described by the tuple $t^*$ with unit roundoff $\acute{\urd} < \urd_{high}$, then the initial change in precision must also be accounted for.
Using~\cref{lemma:dbea}, we can obtain
\begin{align*}
    \tilde{z} = a \times (b + \Delta b) + c + \Delta c,\quad \abs{\Delta b} \leq \gamma_2^{\urd_{low}}; \abs{\Delta c} \leq \urd_{high} \abs{c},
\end{align*}
which is the expression that is then computed using MPFMA.
Using~\cref{lemma:dbea_mpfma_wo_rep}, the computed approximation then satisfies
\begin{align*}
    \hat{z} &= a \times (b + \Delta b + \Delta (b + \Delta b)) + (c + \Delta c) + \Delta (c + \Delta c),\\
            &\define a \times b + c + \Delta p,
\end{align*}
where $\abs{\Delta (b + \Delta b)} \leq (\bar{\urd} + \urd_{FMA} + \bar{\urd}\urd_{FMA})\abs{b + \Delta b}$, $\abs{\Delta (c + \Delta c)} \leq (\bar{\urd} + \urd_{FMA} + \bar{\urd}\urd_{FMA})\abs{c + \Delta c}$, and $\Delta p \define a\Delta b + a\Delta(b + \Delta b) + \Delta c + \Delta (c + \Delta c)$.
We can then bound $\Delta p$ to obtain the forward error bound, as stated in~\cref{lemma:dbea_mpfma_w_rep}.
\begin{lemma}[MPFMA with representation error]
    \label{lemma:dbea_mpfma_w_rep}
    Let the multiply-accumulate operation $z = a\times b + c$ be evaluated using MPFMA operation, where $a,b$ and $c$ are characterized by the tuple $t^*$ with unit roundoff $\acute{\urd} < \urd_{high}$.
    Consider that the multiply and accumulate operation is performed in a floating point system characterized by $\tilde{t}$ (with unit roundoff $\tilde{u}$) and $\bar{t}$ (with unit roundoff $\bar{u}$), respectively.
    Let the output unit roundoff be given as $\urd_{FMA} > \bar{\urd}$.
    Then, under the IEEE arithmetic model and~\cref{lemma:dbea}, the computed approximation $\hat{z}$ to $z$ satisfies
    \begin{align*}
        \frac{\abs{\hat{z} - z}}{\abs{z}} \leq \frac{\big[ \gamma_2^{\urd_{low}} + \zeta (1 + \gamma_2^{\urd_{low}})\big] \abs{a}\abs{b} + \big[ \urd_{high} + \zeta (1 + \urd_{high})\big]\abs{c}}{\abs{a \times b + c}},
    \end{align*}
    where
    $\zeta \define \bar{\urd} + \urd_{FMA} + \bar{\urd}\urd_{FMA}$.
\end{lemma}

\subsubsection{Probabilistic Analysis}\label{sec:vibea_mpfma}
It is trivial to show that VIBEA results in the same backward and forward error bounds for MPFMA without any representation error, which always holds.
In the case there is precision change, we can use~\cref{lemma:vibea} to obtain
\begin{align*}
    \tilde{z} = a\times (b + \Delta b) + (c + \Delta c),
\end{align*}
where $\abs{\Delta b} \leq \tilde{\gamma}_2^{\urd_{low}}$ holds with probability $\prob[b]{\lambda, \urd_{low}, 2}$ and $\abs{\Delta c}\leq \urd_{high} \abs{c}$.
Following~\cref{sec:dbea_mpfma}, the probabilistic backward error bounds can be obtained as stated in~\cref{lemma:vibea_mpfma_w_rep}.
\begin{lemma}[MPFMA with representation error]
    \label{lemma:vibea_mpfma_w_rep}
    Let the multiply-accumulate operation $z = a\times b + c$ be evaluated using MPFMA operation, where $a,b$ and $c$ are characterized by the tuple $t^*$ with unit roundoff $\acute{\urd} < \urd_{high}$.
    Consider that the multiply and accumulate operation is performed in a floating point system characterized by $\tilde{t}$ (with unit roundoff $\tilde{u}$) and $\bar{t}$ (with unit roundoff $\bar{u}$), respectively.
    Let the output unit roundoff be given as $\urd_{FMA} > \bar{\urd}$.
    Then, under the IEEE arithmetic model and~\cref{lemma:vibea}, the computed approximation $\hat{z}$ to $z$ satisfies
    \begin{align*}
        \frac{\abs{\hat{z} - z}}{\abs{z}} \leq \frac{\big[ \tilde{\gamma}_2^{\urd_{low}} + \zeta (1 + \tilde{\gamma}_2^{\urd_{low}})\big] \abs{a}\abs{b} + \big[ \urd_{high} + \zeta (1 + \urd_{high})\big]\abs{c}}{\abs{a \times b + c}},
    \end{align*}
    with a probability at least $\prob[b]{ \lambda, \urd_{low}, 2 }$, where $\zeta \define \bar{\urd} + \urd_{FMA} + \bar{\urd}\urd_{FMA}$.
\end{lemma}

\subsection{Tensor Cores (TCs)}\label{sec:tensor_cores}
\begin{figure}[h]
    \centering
    \includegraphics[width=0.8\textwidth]{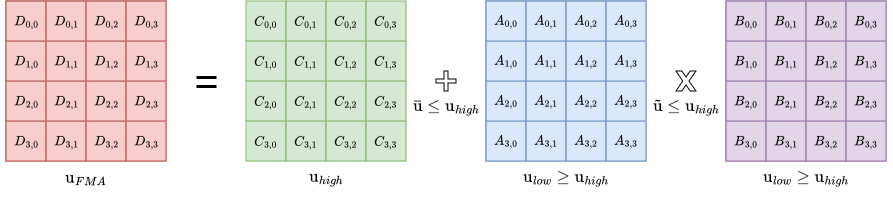}
    \caption{Schematic for Tensor core matrix multiply and accumulate.}
    \label{fig:tc_block}
\end{figure}

Tensor cores were introduced in the NVIDIA Volta architecture as dedicated hardware to perform MPFMA on matrices.
They are designed to perform the operation $\mat{D} = \mat{C} + \mat{AB}$, where $\mat{C}\ireal{b_1\times b_2}$, $\mat{A}\ireal{b_1\times b}$, and $\mat{B}\ireal{b\times b_2}$ in a single clock cycle for $b_1=b_2=b=4$.
To perform MPFMA, TCs take the input matrix $\mat{A}, \mat{B}$ in a lower precision (with unit roundoff $\urd_{low}$) and the accumulator $\mat{C}$ in a higher precision (with unit roundoff $\urd_{high}\leq \urd_{low}$).
Each TC performs 64 MPFMA operations (16 dot products of vector of size 4) with a multiply operation in a floating point system $\tilde{t}$ (with unit roundoff $\tilde{\urd}$) and an accumulation in the system $\bar{t}$ (with unit roundoff $\bar{\urd}$), as illustrated in \cref{fig:tc_block}.
The NVIDIA programming guide states
\begin{quote}
\textit{
Element-wise multiplication of matrix $\mat{A}$ and $\mat{B}$ is performed with at least single precision.
When .ctype or .dtype is .f32, accumulation of the intermediate values is performed with at least a single precision.
When both .ctype and .dtype are specified as .f16, the accumulation is performed with at least half precision.
}
\end{quote}
To this end, we summarize the unit roundoff utilized by TCs in~\cref{tab:tc_params}.
As discussed in~\cref{sec:mpfma}, the choice of $\tilde{\urd}$ ensures that multiplication does not introduce any error.
This means that while the present architecture is designed to handle matrices $\mat{A}$ and $\mat{B}$ in $\float{16}$, there is a scope for further quantization.
TCs are particularly useful as they can efficiently multiply matrix-matrix by decomposing the output matrix into smaller blocks, as described in~\cref{alg:mm_tc}.

\begin{algorithm}
    \caption{Let $\mat{D} = \mat{A} \mat{B}$, where $\mat{A}\ireal{m\times n}$ and $\mat{B}\ireal{n \times t}$ be computed using Tensor cores by partitioning the output matrix $\mat{D}$ into smaller blocks of size $b_1 \times b_2$ (denoted as $D_{ij}$).
    Assume that $m\define pb_1$, $n\define q b$, and $t\define r b_2$, where $p, q, r\in\mathbb{Z}_{\geq1}$.
    The block $\mat{D}_{ij}\define \sum_{k=1}^{q} \mat{A}_{ik} \mat{B}_{k j}$, where $\mat{A}_{ik}$ and $\mat{B}_{kj}$ denotes a block of $\mat{A}$ and $\mat{B}$ of size $b_1\times b$ and $b \times b_2$, respectively.
    Consider that $\mat{A},\mat{B}$ are stored in the floating point system $t_{low}$ with unit roundoff $\urd_{low}$.
    Consider that the multiplication and accumulation are performed in the floating point system $\tilde{t}$ (with unit roundoff $\tilde{\urd}$) and $\bar{t}$ (with unit roundoff $\bar{\urd}$), respectively.
    The output unit roundoff is $\urd_{FMA}$.
    }
    \label{alg:mm_tc}
\begin{algorithmic}
    \State $\tilde{\mat{A}} \gets fl_{\urd_{low}}(\mat{A})$ and $\tilde{\mat{B}} \gets fl_{\urd_{low}}(\mat{B})$.
    \For{$i = 1$ to $p$}
        \For{$j = 1$ to $r$}
            \State $\mat{D}_{ij} \gets \mathbf{0}$.
            \For{$k = 1$ to $q$}
                \State Compute $\mat{D}_{ij} \gets fl_{\urd_{FMA}} (\mat{D}_{ij} + \tilde{\mat{A}}_{ik} \tilde{\mat{B}}_{kj})$
            \EndFor
        \EndFor
    \EndFor
\end{algorithmic}
\end{algorithm}

\cite{blanchard2020mixed} presented a general framework to perform deterministic backward error analysis for~\cref{alg:mm_tc}.
In this work, we first consider a special case wherein the choice of $\tilde{t}$ ensures that the multiply does not introduce any rounding error.
% As discussed in~\cref{sec:mpfma}, this assumption is critical to ensure that the multiplication operation does not introduces any addition error.
In this section, under this assumption, we first re-derive the deterministic analysis of~\cite{blanchard2020mixed}.
Following this, we present a probabilistic backward error analysis for TCs.

\begin{table}[ht]
\centering
\begin{tabular}{c c c c c c c c c}
\toprule
    $\mat{A}$ & $\mat{B}$ & $\mat{C}$ & $\mat{D}$ & $\urd_{low}$ & $\urd_{high}$ & $\tilde{\urd}$ & $\bar{\urd}$ & $\urd_{FMA}$\\
\midrule
    \float{16} & \float{16} & \float{32} & \float{32} & $\urd_{\float{16}}$ & $\urd_{\float{32}}$ & $\urd_{\float{32}}$ & $\urd_{\float{32}}$ & $\urd_{\float{32}}$\\
    \float{16} & \float{16} & \float{16} & \float{16} & $\urd_{\float{16}}$ & $\urd_{\float{16}}$ & $\urd_{\float{32}}$ & $\urd_{\float{16}}$ & $\urd_{\float{16}}$\\
\bottomrule
\end{tabular}
\caption{Unit roundoff for Tensor Cores}
\label{tab:tc_params}
\end{table}

\subsubsection{Deterministic Analysis}
\begin{figure}[h]
    \centering
    \includegraphics[width=0.5\textwidth]{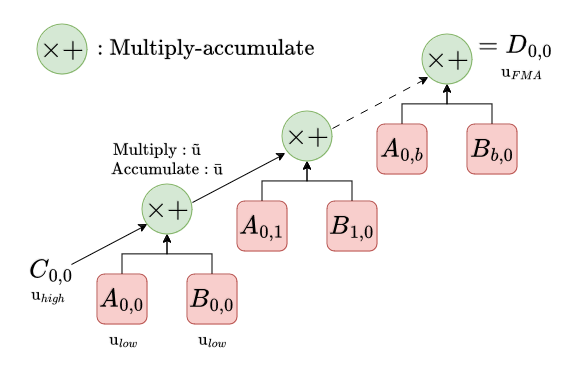}
    \caption{An illustration of the Tensor core kernel.}
    \label{fig:tc_kernel}
\end{figure}

Tensor cores recursively utilize the kernel $z_i = z_{i-1} + \sum_{j=1}^{b} x_j y_j$, where $x_i, y_j$ are given in the system $t_{low}$ and $z_{i-1}$ is given in the system $t_{high}$.
Multiplications and accumulations are performed in $\tilde{t}$ and $\bar{t}$, respectively; output $z_i$ is in the floating point system $t_{FMA}$.
An illustration of how this kernel is utilized is shown in~\cref{fig:tc_kernel}.
Under the assumption that performing the multiplication in $\tilde{t}$ does not introduce any additional error, we can use~\cref{lemma:dbea} to obtain the computed approximation
\begin{align}
    \label{eqn:tc_kernel}
    \hat{z}_i = \Big(z_{i-1}\prod_{k=1}^b (1+\delta_k) + \sum_{j=1}^{b} x_j y_j \prod_{k=j}^{b} (1 + \delta_k) \Big ) (1+\eta_i);\quad \abs{\delta_k} \leq \bar{\urd}, \abs{\eta_i} \leq \urd_{FMA},
\end{align}
assuming left-to-right computation.
If $\vect{x}$ denotes a row of $\tilde{\mat{A}}$ and $\vect{y}$ denotes the $\tilde{\mat{B}}$, then using~\cref{alg:mm_tc} an element of the output matrix $\mat{D}$ is given as
\begin{align*}
    s_n &\define \underbrace{\vect{x}^T \vect{y}}_{\text{Computed in blocks of size $b$}},\\
    &= \sum_{i=1}^{q}\sum_{j=(i-1)b + 1}^{ib} x_{j} y_{j},\\
\end{align*}
Using~\cref{eqn:tc_kernel}, the computed approximation $\hat{s}_n$ is given as
\begin{align*}
    \hat{s}_n = \sum_{i=1}^{q} \Biggl(\sum_{j=(i-1)b + 1}^{ib} x_j y_j \prod_{k = ((j-1)\bmod b) + 1}^{b}(1+\delta_k^{(i)}) \Biggr)\overbrace{\prod_{\ell=\max(2, i)}^{q} \prod_{\kappa=1}^{b} (1 + \delta_\kappa^{(\ell)})}^{\text{Adding previous block}}\underbrace{\prod_{\varrho=i}^{q} (1 + \eta_\varrho)}_{\text{Conversion to }t_{FMA}},
\end{align*}
where $\abs{\delta_k^{(i)}}, \abs{\delta_\kappa^{(\ell)}} \leq \bar{\urd}$, $\abs{\eta_\ell} \leq \urd_{FMA}$, and $\delta_{k=1}^{i=1} = 0$ (adding to $0$ does not introduce any error).
Using~\cref{lemma:dbea} and~\cite[Lemma 3.3]{higham2002accuracy}
\begin{align*}
    \hat{s}_n &= \sum_{i=1}^{q} \Biggl(\sum_{j=(i-1)b + 1}^{ib} x_j y_j (1 + \theta_{b - (j-1)\bmod b}^{\bar{\urd}(i)}) \Biggr)(1 + \theta_{b*(q - \max(2, i) + 1)}^{\bar{\urd}})(1 + \theta_{q - i + 1}^{\urd_{FMA}}),\\
              &= \sum_{i=1}^{q} \sum_{j=(i-1)b + 1}^{ib} x_j y_j (1 + \theta_{b - ((j-1)\bmod b) + b* (q - \max(2, i) + 1)}^{\bar{\urd}(i)})(1 + \theta_{q - i + 1}^{\urd_{FMA}}),\\
\end{align*}
where $\abs{\theta_{\zeta}^{\bar{\urd}}} \leq \gamma_{\zeta}^{\bar{\urd}}$ and $\abs{\theta_{\zeta}^{\urd_{FMA}}} \leq \gamma_{\zeta}^{\urd_{FMA}}$.
Thus,
\begin{align*}
    \hat{s}_n = \sum_{i=1}^{q} \sum_{j=(i-1)b + 1}^{ib} x_j (y_j + \Delta y_j),
\end{align*}
where $\Delta y_j \define (\theta_{b-((j-1)\bmod b) + b * (q - \max(2, i) + 1)}^{\bar{\urd}(i)} + \theta_{q - i + 1}^{\urd_{FMA}} + \theta_{b-((j-1)\bmod b) + b * (q - \max(2, i) + 1)}^{\bar{\urd}(i)} * \theta_{q - i + 1}^{\urd_{FMA}})y_j$, such that
\begin{align*}
    \abs{\Delta y_j} \leq \abs{\theta_{b-((j-1)\bmod b) + b * (q - \max(2, i) + 1)}^{\bar{\urd}(i)}} + \abs{\theta_{q - i + 1}^{\urd_{FMA}}} + \abs{\theta_{b-((j-1)\bmod b) + b * (q - \max(2, i) + 1)}^{\bar{\urd}(i)}} \abs{\theta_{q - i + 1}^{\urd_{FMA}})}\abs{y_j}
\end{align*}
which can be upper bounded for $i=j=1$ (for which $\theta_\zeta^{\bar{\urd}}$ is effectively $\theta_{\zeta-1}^{\bar{\urd}}$ since $\delta_{j=1}^{i=1} = 0$) as
\begin{align*}
    \abs{\Delta y_j} &\leq (\gamma_{n - 1}^{\bar{\urd}} + \gamma_{q}^{\urd_{FMA}} + \gamma_{n - 1}^{\bar{\urd}} *\gamma_{q}^{\urd_{FMA}})\abs{y_j}.
\end{align*}
This gives the forward error in computing an element of $\mat{d}$ as
\begin{align*}
    \frac{\abs{\hat{s}_n - s_n}}{\abs{s_n}} \leq (\gamma_{n - 1}^{\bar{\urd}} + \gamma_{q}^{\urd_{FMA}} + \gamma_{n - 1}^{\bar{\urd}} *\gamma_{q}^{\urd_{FMA}}) \frac{\abs{\vect{x}}^T\abs{\vect{y}}}{\abs{\vect{x}^T \vect{y}}}
\end{align*}
Compared to~\cite{blanchard2020mixed}, we obtain a factor of $n-1$ due to the assumption of errorless multiplication in the floating point system $\tilde{t}$.

Using this result, we can obtain the forward error for performing the matrix-matrix multiplication, as stated in~\cref{theorem:dbea_tc_wo_rep}.
\begin{theorem}[Tensor cores without representation error]\label{theorem:dbea_tc_wo_rep}
    Let $\mat{d} = \mat{a}\mat{b}$ be computed using Tensor Cores, where $\mat{a}, \mat{b}$ are given in the floating point system $t_{low}$.
    If~\cref{alg:mm_tc} is used to compute the matrix-matrix multiplication, then under~\cref{lemma:dbea}, the computed computed approximation $\hat{\mat{d}}$ satisfies
    \begin{align*}
        \frac{\abs{\hat{\mat{d}} - \mat{d}}}{\abs{\mat{d}}} \leq (\gamma_{n - 1}^{\bar{\urd}} + \gamma_{q}^{\urd_{FMA}} + \gamma_{n - 1}^{\bar{\urd}} *\gamma_{q}^{\urd_{FMA}}) \frac{\abs{\mat{a}}\abs{\mat{b}}}{\abs{\mat{a}\mat{b}}}.
        % \hat{\mat{d}} = \mat{a} (\mat{b} + \Delta \mat{b}),
    \end{align*}
    % where $\abs{\Delta \mat{b}} \leq (\gamma_{n - 1}^{\bar{\urd}} + \gamma_{q}^{\urd_{FMA}} + \gamma_{n - 1}^{\bar{\urd}} *\gamma_{q}^{\urd_{FMA}})\abs{\mat{b}}$.
\end{theorem}

In the case where $\mat{a}$ and $\mat{b}$ are given in a floating point system $t^*$ with unit roundoff $\acute{\urd} < \urd_{low}$, then we must account for the initial change in precision.
Using~\cref{lemma:dbea} we can obtain
\begin{align*}
    \tilde{\mat{d}} = (\mat{a} + \Delta \mat{a}) (\mat{b} + \Delta \mat{b}),\quad \abs{\Delta \mat{a}} \leq \urd_{low} \abs{\mat{a}}, \abs{\Delta \mat{b}} \leq \urd_{low} \abs{\mat{b}},
\end{align*}
which is then computed using~\cref{alg:mm_tc}.
This gives the computed approximation to $\mat{d}$ as
\begin{align*}
    \hat{\mat{d}} &= (\mat{a} + \Delta \mat{a}) (\mat{b} + \Delta \mat{b}) + \Delta \mat{p};\quad \abs{\Delta \mat{p}} \leq (\gamma_{n - 1}^{\bar{\urd}} + \gamma_{q}^{\urd_{FMA}} + \gamma_{n - 1}^{\bar{\urd}} *\gamma_{q}^{\urd_{FMA}})\abs{\mat{a} + \Delta \mat{a}}\abs{\mat{b} + \Delta \mat{b}},\\
                  &= \mat{a}\mat{b} + \mat{a}\Delta\mat{b} +  \Delta\mat{a}\mat{b} + \Delta\mat{a}\Delta\mat{b} +  \Delta \mat{p}\\
                  &\define \mat{a} \mat{b} + \Delta \mat{e},
    % \hat{\mat{d}} &=  (\mat{a} + \Delta \mat{a}) (\mat{b} + \Delta \mat{b} + \Delta (\mat{b} + \Delta\mat{b})); \quad \abs{\Delta (\mat{b} + \Delta \mat{b})} \leq (\gamma_{n - 1}^{\bar{\urd}} + \gamma_{q}^{\urd_{FMA}} + \gamma_{n - 1}^{\bar{\urd}}\gamma_{q}^{\urd_{FMA}}) \abs{\mat{b} + \Delta \mat{b}},\\
    %               &\define \mat{a} \mat{b} + \Delta \mat{p},
\end{align*}
where $\Delta \mat{e} \define \mat{a}\Delta\mat{b} +  \Delta\mat{a}\mat{b} + \Delta\mat{a}\Delta\mat{b} +  \Delta \mat{p}$ which we can bound to obtain the forward error bounds.
This is stated in~\cref{theorem:dbea_tc_w_rep}.
\begin{theorem}[Tensor cores with representation error]\label{theorem:dbea_tc_w_rep}
    Let $\mat{d} = \mat{a}\mat{b}$ be computed using Tensor Cores, where $\mat{a}, \mat{b}$ are given in the floating point system $t^*$ with unit roundoff $\acute{\urd} < \urd_{low}$.
    If~\cref{alg:mm_tc} is used to compute the matrix-matrix multiplication, then under~\cref{lemma:dbea}, the computed computed approximation $\hat{\mat{d}}$ satisfies
    \begin{align*}
        \frac{\abs{\hat{\mat{d}} - \mat{d}}}{\abs{\mat{d}}} \leq (2\urd_{low} + \urd^2 + \zeta(1 + \urd_{low})^2)\frac{\abs{\mat{a}}\abs{\mat{b}}}{\abs{\mat{a}\mat{b}}},
    \end{align*}
    where $\zeta\define(\gamma_{n - 1}^{\bar{\urd}} + \gamma_{q}^{\urd_{FMA}} + \gamma_{n - 1}^{\bar{\urd}} *\gamma_{q}^{\urd_{FMA}})$.
\end{theorem}

\subsubsection{Probabilistic Analysis}
Consider the computed $\hat{s}_n$ as
\begin{align*}
    \hat{s}_n &= \sum_{i=1}^{q} \Biggl(\sum_{j=(i-1)b + 1}^{ib} x_j y_j \prod_{k = ((j-1)\bmod b) + 1}^{b}(1+\delta_k^{(i)}) \Biggr)\prod_{\ell=\max(2, i)}^{q} \prod_{\kappa=1}^{b} (1 + \delta_\kappa^{(\ell)})\prod_{\varrho=i}^{q} (1 + \eta_\varrho),\\
              &= \sum_{i=1}^{q} \Biggl(\sum_{j=(i-1)b + 1}^{ib} x_j y_j \prod_{k = ((j-1)\bmod b) + 1}^{b}(1+\delta_k^{(i)})\prod_{\ell=\max(2, i)}^{q} \prod_{\kappa=1}^{b} (1 + \delta_\kappa^{(\ell)}) \Biggr)\prod_{\varrho=i}^{q} (1 + \eta_\varrho),\\
              &= \sum_{i=1}^{q} \Biggl(\sum_{j=(i-1)b + 1}^{ib} x_j y_j (1 + \tilde{\theta}_{b - \{(j-1) \bmod b\} + b*(q-\max(2, i) + 1)}^{\bar{\urd}(i)}) \Biggr)(1 + \tilde{\theta}_{q-i + 1}^{\urd_{FMA}}),
\end{align*}
where using~\cref{lemma:vibea}, $\abs{\tilde{\theta}_\zeta^{\bar{\urd}}} \leq \tilde{\gamma}_\zeta^{\bar{\urd}}$ with probability at least $\prob[b]{ \lambda, \bar{\urd}, \zeta }$ and $\tilde{\theta}_{\zeta}^{\urd_{FMA}} \leq \abs{\tilde{\gamma}_{\zeta}}$ with probability at least $\prob[b]{\lambda, \urd_{FMA}, \zeta}$.
Thus,
\begin{align*}
    \hat{s}_n = \sum_{i=1}^{q} \sum_{j=(i-1)b + 1}^{ib} x_j (y_j + \Delta y_j),
\end{align*}
where $\Delta y_j \define (\theta_{b-((j-1)\bmod b) + b * (q - \max(2, i) + 1)}^{\bar{\urd}(i)} + \theta_{q - i + 1}^{\urd_{FMA}} + \theta_{b-((j-1)\bmod b) + b * (q - \max(2, i) + 1)}^{\bar{\urd}(i)} * \theta_{q - i + 1}^{\urd_{FMA}})y_j$, such that
\begin{align*}
    \abs{\Delta y_j} \leq \abs{\theta_{b-((j-1)\bmod b) + b * (q - \max(2, i) + 1)}^{\bar{\urd}(i)}} + \abs{\theta_{q - i + 1}^{\urd_{FMA}}} + \abs{\theta_{b-((j-1)\bmod b) + b * (q - \max(2, i) + 1)}^{\bar{\urd}(i)}} \abs{\theta_{q - i + 1}^{\urd_{FMA}})}\abs{y_j},
\end{align*}
holds with probability at least $1 - \{(1 - \prob[b]{ \lambda, \bar{\urd}, c_1 }) + (1 - \prob[b]{\lambda, \urd_{FMA}, c_2})\}$, where $c_1\define b-((j-1)\bmod b) + b * (q - \max(2, i) + 1)$ and $c_2 \define q - i + 1$.
We can obtain the upper bound for $i=j=1$ (for which $\theta_\zeta^{\bar{\urd}}$ is effectively $\theta_{\zeta-1}^{\bar{\urd}}$ since $\delta_{j=1}^{i=1} = 0$)
then relax the bounds to obtain $\abs{\tilde{\theta}_{b - \{(j-1) \bmod b\} + b*(q-\max(2, i) + 1)}^{\bar{\urd}}} \leq \tilde{\gamma}_{n-1}^{\bar{\urd}}$ and $\abs{\tilde{\theta}_{q-i+1}} \leq \tilde{\gamma}_{q}^{\urd_{FMA}}$.
Thus,
\begin{align*}
    \abs{\Delta y_j} &\leq (\gamma_{n - 1}^{\bar{\urd}} + \gamma_{q}^{\urd_{FMA}} + \gamma_{n - 1}^{\bar{\urd}} *\gamma_{q}^{\urd_{FMA}})\abs{y_j}.
\end{align*}
The forward error bound for matrix-matrix multiplication is then given as stated in~\cref{theorem:vibea_tc_wo_rep}.
\begin{theorem}[Tensor cores without representation error]\label{theorem:vibea_tc_wo_rep}
    Let $\mat{d} = \mat{a}\mat{b}$ be computed using Tensor Cores, where $\mat{a}, \mat{b}$ are given in the floating point system $t_{low}$.
    If~\cref{alg:mm_tc} is used to compute the matrix-matrix multiplication, then under~\cref{lemma:vibea}, the computed computed approximation $\hat{\mat{d}}$ satisfies
    \begin{align*}
        \frac{\abs{\hat{\mat{d}} - \mat{d}}}{\abs{\mat{d}}} \leq (\gamma_{n - 1}^{\bar{\urd}} + \gamma_{q}^{\urd_{FMA}} + \gamma_{n - 1}^{\bar{\urd}} *\gamma_{q}^{\urd_{FMA}}) \frac{\abs{\mat{a}}\abs{\mat{b}}}{\abs{\mat{ab}}}
    \end{align*}
    holds with a probability of at least
    \begin{align*}
         1 - m*t\Big(\sum_{i=1}^{q}\sum_{j=(i-1)b + 1}^{ib}2 - \prob[b]{\lambda, \bar{\urd}, c_1} - \prob[b]{\lambda, \urd_{FMA}, c_2}\Big),
    \end{align*}
    where $c_1\define b-((j-1)\bmod b) + b * (q - \max(2, i) + 1)$ and $c_2 \define q - i + 1$.
\end{theorem}

In the case where $\mat{a}$ and $\mat{b}$ are given in a floating point system $t^*$ with unit roundoff $\acute{\urd} < \urd_{low}$, then we must account for the initial change in precision.
Using~\cref{lemma:vibea} we can obtain
\begin{align*}
    \tilde{\mat{d}} = (\mat{a} + \Delta \mat{a}) (\mat{b} + \Delta \mat{b}),\quad \abs{\Delta \mat{a}} \leq \urd_{low} \abs{\mat{a}}, \abs{\Delta \mat{b}} \leq \urd_{low} \abs{\mat{b}},
\end{align*}
which always holds.
This gives the computed approximation to $\mat{d}$ as
\begin{align*}
    \hat{\mat{d}} &= (\mat{a} + \Delta\mat{a})(\mat{b} + \Delta\mat{b}) + \Delta\mat{p};\quad \abs{\Delta\mat{p}} \leq (\tilde{\gamma}_{n - 1}^{\bar{\urd}} + \tilde{\gamma}_{q}^{\urd_{FMA}} + \tilde{\gamma}_{n - 1}^{\bar{\urd}} *\tilde{\gamma}_{q}^{\urd_{FMA}})\abs{\mat{a} + \Delta\mat{a}}\abs{\mat{b} + \Delta\mat{b}},\\
                  &= \mat{a}\mat{b} + \mat{a}\Delta\mat{b} +  \Delta\mat{a}\mat{b} + \Delta\mat{a}\Delta\mat{b} +  \Delta \mat{p}\\
                  &\define \mat{a} \mat{b} + \Delta \mat{e},
\end{align*}
where $\Delta \mat{p} \define \mat{a} \Delta \mat{b} + \mat{a} \Delta (\mat{b} + \Delta \mat{b}) + \Delta \mat{a} \mat{b} +  \Delta\mat{a} \Delta \mat{b} + \Delta \mat{a}\Delta (\mat{b} + \Delta\mat{b})$ which we can bound to obtain the forward error bounds.
This is stated in~\cref{theorem:vibea_tc_w_rep}.
\begin{theorem}[Tensor cores with representation error]\label{theorem:vibea_tc_w_rep}
    Let $\mat{d} = \mat{a}\mat{b}$ be computed using Tensor Cores, where $\mat{a}, \mat{b}$ are given in the floating point system $t^*$ with unit roundoff $\acute{\urd} < \urd_{low}$.
    If~\cref{alg:mm_tc} is used to compute the matrix-matrix multiplication, then under~\cref{lemma:dbea}, the computed computed approximation $\hat{\mat{d}}$ satisfies
    \begin{align*}
        \frac{\abs{\hat{\mat{d}} - \mat{d}}}{\abs{\mat{d}}} \leq (2\urd_{low} + \urd^2 + \zeta(1 + \urd_{low})^2)\frac{\abs{\mat{a}}\abs{\mat{b}}}{\abs{\mat{a}\mat{b}}},
    \end{align*}
     holds with probability at least
    \begin{align*}
         1 - m*t\Big(\sum_{i=1}^{q}\sum_{j=(i-1)b + 1}^{ib}2 - \prob[b]{\lambda, \bar{\urd}, c_1} - \prob[b]{\lambda, \urd_{FMA}, c_2}\Big),
    \end{align*}
    where $c_1\define b-((j-1)\bmod b) + b * (q - \max(2, i) + 1)$, $c_2 \define q - i + 1$, and $\zeta\define(\tilde{\gamma}_{n - 1}^{\bar{\urd}} + \tilde{\gamma}_{q}^{\urd_{FMA}} + \tilde{\gamma}_{n - 1}^{\bar{\urd}} *\tilde{\gamma}_{q}^{\urd_{FMA}})$.

\end{theorem}

\section{Numerical Experiments}\label{sec:results}
In this section, we present the numerical experiments conducted on several kernels.
First, we present the backward error bounds obtained in the case of the Multiply-Accumulate (MAC) unit.
This kernel is fundamental to several kernels, such as dot-products and matrix-matrix products.
We present the error bounds for performing the MAC operation using FMA and MPFMA operations.
Following this, we present the forward error bounds in the case of matrix-matrix multiplication using TCs.
For all error evaluations, computations using \float{64} are considered as the actual result.
\subsection{Multiply-Accumulate (MAC) Unit}\label{sec:results_mac}
\begin{figure}[h]
    \centering
    \includegraphics[scale=0.4]{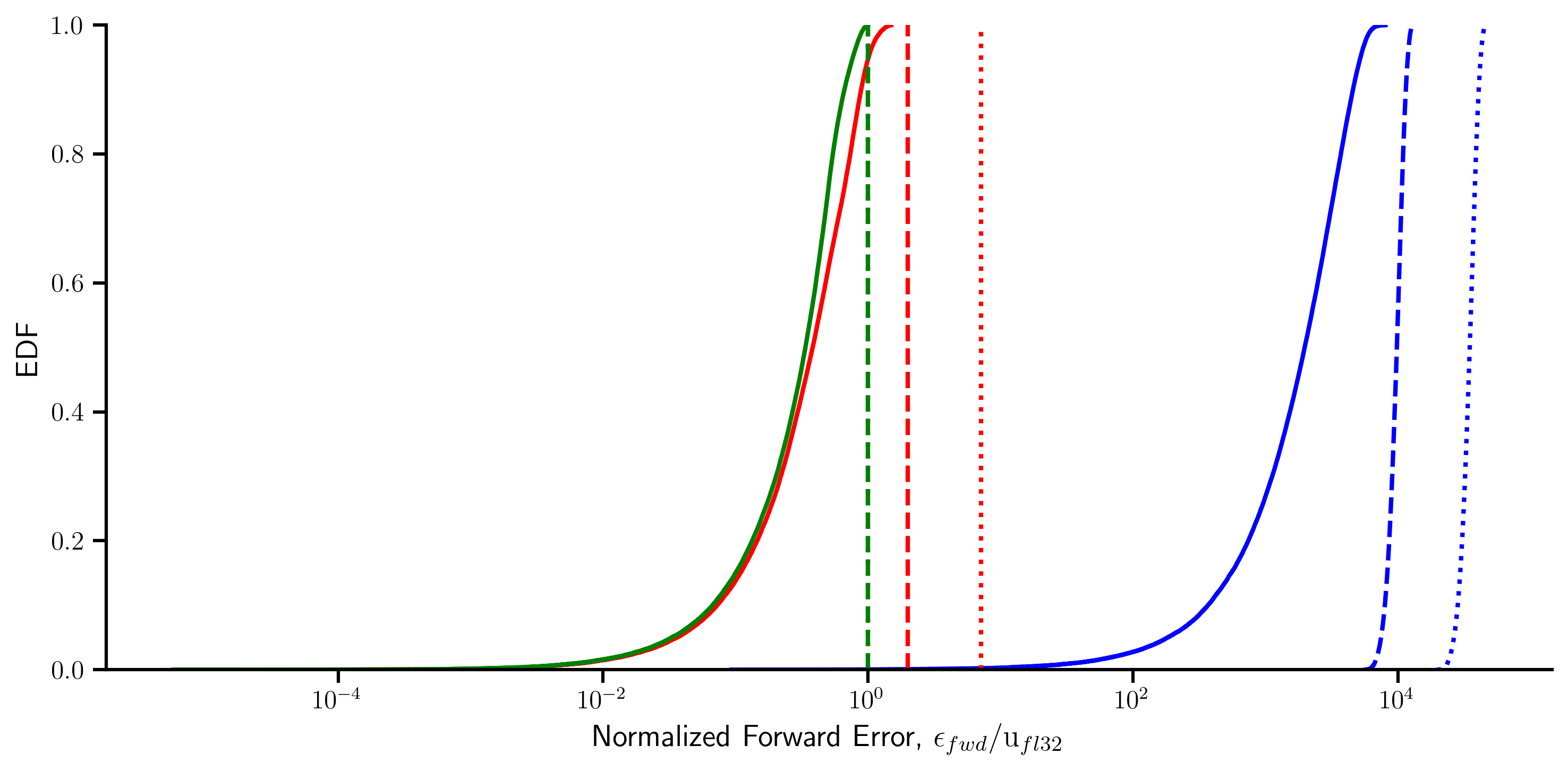}
    \caption{Empirical distribution for the forward error ($\frac{\abs{\hat{d} - d}}{\abs{d}}$) and its bounds for the MAC operation.
    True forward error for the MAC operation computed without FMA operation (\textcolor{red}{\fline}), with FMA operation (\textcolor{green2}{\fline}), and with MPFMA operation (\textcolor{blue}{\fline}).
    Deterministic analysis (DBEA) backward error bounds computed without FMA operation (\textcolor{red}{\dashed}), with FMA operation (\textcolor{green2}{\dashed}), and with MPFMA operation (\textcolor{blue}{\dashed}).
    Probabilistic analysis (VIBEA) backward error bounds computed without FMA operation (\textcolor{red}{\dotted}), with FMA operation (\textcolor{green2}{\dotted}), and with MPFMA operation (\textcolor{blue}{\dotted}).
    DBEA results in tighter estimates than VIBEA in the case of without FMA and MPFMA.
    Both DBEA and VIBEA estimate the backward error for the FMA operation tightly.
    }
    \label{fig:mac_fea}
\end{figure}

Consider the MAC operation $d = a \times b + c$, for which the computed approximation is given as $\hat{d}$.
For $a, b$ and $c$ distributed uniformly as $\mathcal{U}[1, 2]$, the empirical distribution function (EDF) of the forward error and its bounds are shown in~\cref{fig:mac_fea}.
For a random variables $X$, the EDF is defined as $F_{X}(t) \define \frac{1}{n} \sum_{i=1}^{n} \mathbf{1}_{X=x_i \leq t} \in[0, 1]$, where $x_i$ denotes an observation of the random variable $X$ and $\mathbf{1}$ is an indicator function.
Thus, if $F_{X_1}(t) < F_{X_2}(t)$, where $X_1$ and $X_2$ are random variables defined on the same support would mean that there is a larger probability that $X_1$ is less than $t$ compared to $X_2$.
FMA operations result in smaller accumulated rounding errors, as observed by a smaller largest $\epsilon_{fwd}$ compared to the MAC operation without FMA operation.
Mixed-precision introduces a larger backward error due to the initial change in representation from $\float{32}$ to $\float{16}$ for $a$ and $b$.
The forward error bounds from the deterministic analysis are tighter (estimates the largest $\epsilon_{fwd}$ well) compared to the probabilistic ones where MAC is computed without FMA and using MPFMA.
This is expected for a small number of operations per floating point value, as shown in~\cite{bhola2024variance}.
For FMA operations, as shown in~\cref{sec:fma}, DBEA and VIBEA result in the same bounds that tightly estimate the accumulated error.
Despite the larger accumulated error MPFMA, the low precision arithmetic results in smaller memory bandwidth consumption and larger arithmetic intensity, as tabulated in~\cref{tab:mac_profile}.

\begin{table}[ht]
\centering
\begin{tabular}{l c c}
\toprule
    \textbf{Algorithm} & \textbf{Arithmetic intensity} [Flop/Byte] & \textbf{Flops/s} \\
\midrule
    MAC (W/o FMA) & 0.00995 & 0.00004\\
    MAC (W/ FMA) & 0.00989 & 0.00004\\
    MAC (W/ MPFMA) & 0.02442 & 0.00008\\
\bottomrule
\end{tabular}
\caption{MAC operation profile}
\label{tab:mac_profile}
\end{table}

\subsection{Matrix-Matrix Multiplication using Tensor Cores} \label{sec:results_tc}
\begin{figure}[h]
    \centering
    \includegraphics[scale=0.4]{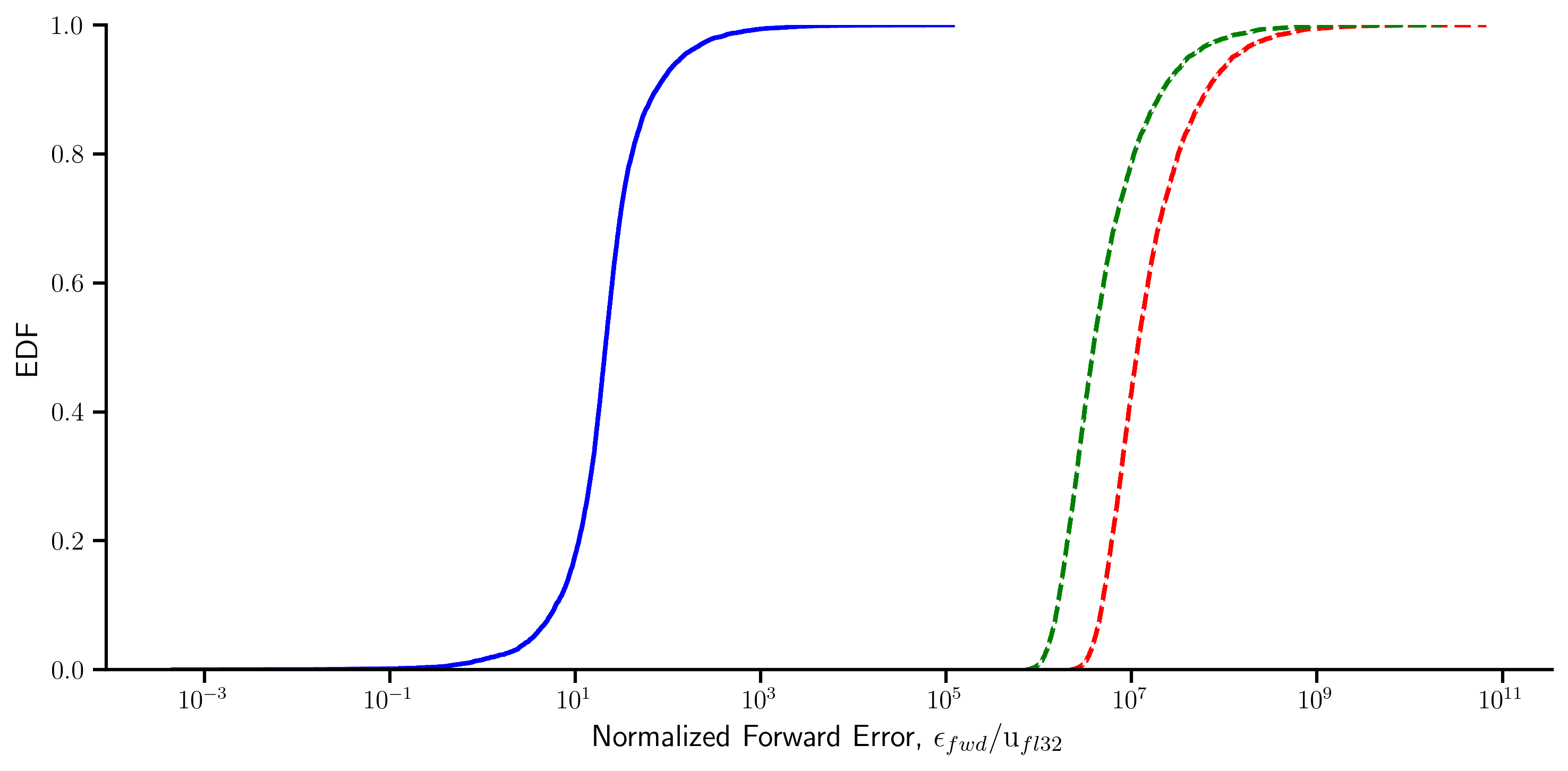}
    \caption{Empirical distribution for the forward error ($\frac{\abs{\hat{d} - d}}{\abs{d}}$) and its bounds for matrix-matrix multiplication ($m=2^{10}, t=2^{15}, n=2^{3}$) using Tensor Cores.
    True forward error for the multiplication (\textcolor{blue}{\fline}).
    Bounds obtained using DBEA (\textcolor{red}{\dashed}) and VIBEA (\textcolor{green2}{\dashed}).
    VIBEA bounds are tighter than DBEA bounds by nearly an order of magnitude.
    }
    \label{fig:matmul_fea}
\end{figure}

Consider the matrix-matrix multiplication $\mat{d} = \mat{a} \mat{b}$, where $\mat{a}\in\mathbb{R}^{m\times t}$, $\mat{b}\in\mathbb{R}^{t\times n}$, and $\mat{d}\in\mathbb{R}^{m\times n}$.
We consider the case where $m=2^{10}$, $t=2^{15}$, and $n=2^{3}$, where the elements of the matrices are drawn from the uniform distribution $\mathcal{U}(-1,1)$.
We use the cuBLAS library on an NVIDIA H100 GPU to compute the matrix-matrix multiplication using Tensor Cores.
The code snippet for the matrix-matrix multiplication is provided in Algorithm~\ref{alg:matmul_tc}.
The empirical distribution of the forward error ($\frac{\abs{\hat{d} - d}}{\abs{d}}$) and its bounds for the matrix-matrix multiplication are shown in Figure~\ref{fig:matmul_fea}.
Storing the matrices in $\float{16}$ and performing MPFMA, the maximum forward error is bounded by $\urd_{\float{32}}\times 10^{11}\approx\mathcal{O}(10^{2})$ using DBEA and $\mathcal{O}(10^{1})$ using VIBEA.
The VIBEA bounds are tighter than the DBEA bounds by nearly an order of magnitude.
However, both DBEA and VIBEA bounds are loose compared to the actual maximum forward error, which is $\mathcal{O}(10^{-2})$.
While DBEA assumes the worst-case scenario, which will always overestimate the error, VIBEA bounds can be improved by considering more informative rounding error distributions.
\newpage

\begin{lstlisting}[language=CUDA, caption={Matrix-matrix multiplication using Tensor Cores.}, label={alg:matmul_tc}]
// cuBLAS handle
cublasHandle_t handle;
cublasCreate(&handle);
int alpha = 1;
int beta = 0;
int lda = m; // Leading dimension of a
int ldb = t; // Leading dimension of b
int ldc = m; // Leading dimension of c
// Compute the matrix-matrix multiplication
cublasGemmEx(handle, CUBLAS_OP_N,CUBLAS_OP_N, m, t, n, &alpha, a, CUDA_R_16F, lda, b, CUDA_R_16F, ldb, &beta, c, CUDA_R_16F, ldc, CUDA_R_32F, CUBLAS_GEMM_DEFAULT_TENSOR_OP);
// Synchronize the device
cudaDeviceSynchronize();
// Destroy the cuBLAS handle
cublasDestroy(handle);
\end{lstlisting}

\section{Concluding Remarks}\label{sec:conclusions}
In this work, we present rounding error analysis for fused multiply-accumulate (FMA), mixed-precision fused multiply-accumulate (MPFMA), and matrix-matrix multiply using Tensor Cores.
These kernels are fundamental to numerical linear algebra and are widely used in scientific computing.
Developing accurate error bounds for these kernels is important for understanding the impact of rounding errors on the accuracy of numerical computations.
To this end, we present deterministic and probabilistic error bounds for these kernels to study the impact of rounding errors on large-scale computations.
Through numerical experiments, we demonstrate that MPFMA results in nearly $\mathcal{O}(10^4)$ times larger error compared to FMA.
However, this error comes at nearly two times increase in performance.
We also show that for large-scale matrix-matrix multiplication TCs can produce reliable computations, with $\mathcal{O}(10^{-2})$ error in the case where data is distributed randomly as $\mathcal{U}(0, 1)$.
Probabilistic error bounds for matrix-matrix multiplication can produce nearly an order or magnitude improvement in estimating rounding errors compared to the deterministic analysis.
However, both probabilistic and deterministic bounds overestimate the maximum error by nearly $\mathcal{O}(10^5)$.

Probabilistic analysis shows promise for estimating rounding error for large-scale computations.
However, there is need to improve the modeling assumptions for rounding errors and using better inductive bias.
For example, future work could consider the leveraging the empirical rounding error distributions proposed in recent work~\cite{dahlqvist2019probabilistic, fang2024probabilistic}.

\section*{Acknowledgments}
This research was supported by the National Science Foundation
grant FMITF-2219997 and by the Los Alamos National Laboratory grant titled `Algorithm/Hardware/Software Codesign for High Energy Density Applications.'
\bibliographystyle{elsarticle-num}
\bibliography{references}
\end{document}